\documentstyle[psfig]{cupconf}

\ifoldfss
\else
  \ifnfssone
    \newmathalphabet{\mathit}
      \addtoversion{normal}{\mathit}{cmr}{m}{it}
      \addtoversion{bold}{\mathit}{cmr}{bx}{it}
    \newmathalphabet{\mathcal}
      \addtoversion{normal}{\mathcal}{cmsy}{m}{n}
    \else
    \ifnfsstwo
    \fi
  \fi
\fi

\def\h2{\nobreak\mbox{$\;$H$_{2}$}}
\def\cm2{\nobreak\mbox{$\;$cm$^{-2}$}}

\def\kkms{\nobreak\mbox{$\;$K\,km\,s$^{-1}$}}
\def\hexnumber#1{\ifcase#1 0\or1\or2\or3\or4\or5\or6\or7\or8\or9\or
 A\or B\or C\or D\or E\or F\fi }

\makeatletter
\ifx\CUP@mtlplain@loaded\undefined
\else
\fi
\makeatother
\makeatletter
\ifx\CUP@mtlplain@loaded\undefined
\else
\fi
\makeatother
\title[Extragalactic H$_{2}$]{Extragalactic H$_{2}$ and its variable relation to CO}

\author[F.P. Israel]
{F.P. Israel}

\affiliation{Sterrewacht Leiden, Postbus 9513, 2300 RA Leiden, the Netherlands}

\begin{document}
\ifnfssone
\else
  \ifnfsstwo
  \else
    \ifoldfss
      \let\mathcal\cal
      \let\mathrm\rm
      \let\mathsf\sf
    \fi
  \fi
\fi

\maketitle

\begin{abstract}
We derive and discuss the strong dependence on metallicity of the CO to $\h2$ 
conversion factor $X\,=\,N(\h2)/I_{CO}\,=\,12.2\,-\,2.5\,log\,[O]/[H]$
appropriate to extragalactic objects, as well as the weaker dependence found 
for such objects from interferometer measurements.
\end{abstract}

\firstsection % if your document starts with a section,
\section{Introduction}

The difficulty of directly observing molecular hydrogen ($\h2$), the major 
constituent of the interstellar medium in galaxies, and ways of doing so 
indirectly are reviewed elsewhere in this volume (Combes 2000). Usually,
$\h2$ cloud properties are derived by extrapolation from more easily 
conducted CO observations. For instance, observed CO cloud sizes and 
velocity widths yield total molecular gas masses under the assumption of
virial equilibrium. However, in extragalactic systems especially, this 
method is beset by pitfalls (see Israel, 1997, hereafter Is97) and requires 
high linear resolutions (i.e. use of interferometer arrays). More 
seriously, {\it the fundamental assumption of virialization appears to 
be false}. As individual components (`clumps') have velocities of only a 
few km s$^{-1}$ and CO complex sizes are 50--100 pc, crossing times are 
comparable to CO complex {\it lifetimes} of only a few times 10$^{7}$ years 
or less (Leisawitz et al. 1989; Fukui et al. 2000; see also Elmegreen 2000). 
As equilibrium cannot be reached in a single crossing time or less, the 
virial theorem is not applicable to such complexes. Indeed, the elongated 
and interconnected filamentary appearance of many large CO cloud complexes 
do not suggest virialized systems (see also Maloney 1990).

The observed CO intensity is the weighted product of CO brightness 
temperature and emitting surface area; actual CO column densities are 
completely hidden by high optical depths. However, in large beams CO 
cloud ensembles may be assumed to be statistically identical so that 
CO intensities scale with CO mass within the beam, i.e. beam-averaged CO 
column density. If we can determine the proportionality, the $\h2$-to-CO 
conversion factor $X$, subsequent CO measurements can be used to find the 
appropriate $\h2$ column density and mass. In the Milky Way, the calibration 
of $X$ is controversial by a factor of about two (cf. Combes 2000), and 
frequently based on application of the virial theorem (but see preceding
paragraph ...).  

In other extragalactic environments, the assumption of statistical
CO cloud ensemble similarity becomes questionable. Very clumpy, even fractal
molecular clouds are very sensitive to e.g. variations in radiation field 
intensity and metallicity. As $\h2$ and CO, supposedly tracing $\h2$, 
react differently to such variations, $X$ is also very sensitive to them 
(Maloney $\&$ Black 1988). The determination of the dependence of $X$ on 
metallicity and radiation field intensity, needed to correctly estimate 
amounts of $\h2$ in environments (dwarf galaxies, galaxy centers) different 
from the Solar Neighbourhood thus requires $\h2$ mass determinations 
independent of CO.

\section{$\h2$ determinations from dust continuum emission}

Fortunately, $\h2$ and HI column densities are traced by optically thin
continuum emission from associated dust particles. Unfortunately, 
dust emissivities depend strongly on temperature, dust particle properties 
are not accurately known and dust-to-gas ratios are frequently uncertain.
The effect of these uncertainties are minimized if we can avoid the need for 
determining absolute values of the dust column density and the dust-to-gas 
ratio. Far-infrared/submillimeter continuum fluxes and HI intensities from
spatially nearby positions, preferably in dwarf galaxies that lack strong 
temperature or metallicity gradients, can be used to obtain reasonably 
accurate $\h2$ column densities (Is97). The ratio of dust continuum emission 
to HI column density at locations lacking substantial molecular gas provides 
{\it a measure} for the dust-to-gas column density ratio. Without requiring
its absolute value, we can apply this measure to a nearby location rich in 
molecular gas to find the total hydrogen column density and, after subtraction
of HI, the $\h2$ column density. Division by the CO intensity yields the local
value of $X$ in absolute units with an accuracy better than a factor of two 
(Is97).

Individual molecular cloud complexes in the nearby Magellanic Clouds were 
used by Is97 to determine the effects of radiation field intensity (as 
sampled by far-infrared surface brightness) on $X$. Over a large range of 
intensities, $X$ is linearly proportional to the radiative energy available 
per nucleon ($\sigma$). Quiescent regions in the LMC yield $X$ values close 
to those of the Solar Neighbourhood, whereas a value 40 times 
higher is obtained for the radiation-saturated 30 Doradus region.
The more metal-poor SMC exhibits higher $X$ values, but again linearly
proportional to $\sigma$. 

\section{Dependence of $X$ on metallicity}

To further study the relation between $X$ and metallicity, we have added
several recent results to the database given by Is97. These include 
data for NGC~7331 (3 points; Israel $\&$ Baas 1999), the Milky Way center
and centers of NGC~253 (both from Dahmen et al. 1998), NGC~891 (Gu\'elin et 
al. 1993; Israel et al. 1999), NGC~3079 (Braine et al. 1997; Israel et al. 
1998a) as well as IC~10 (Madden et al. 1997) and D~478 in M~31 (Israel et al. 
1998b). Although they were obtained somewhat differently from those in  
Is97, they are quite comparable (Figs. 1 and 2). 

\begin{figure}[t]
\unitlength1cm
\begin{minipage}[t]{6.5cm}
\begin{picture}(6.5,9)
\psfig{figure=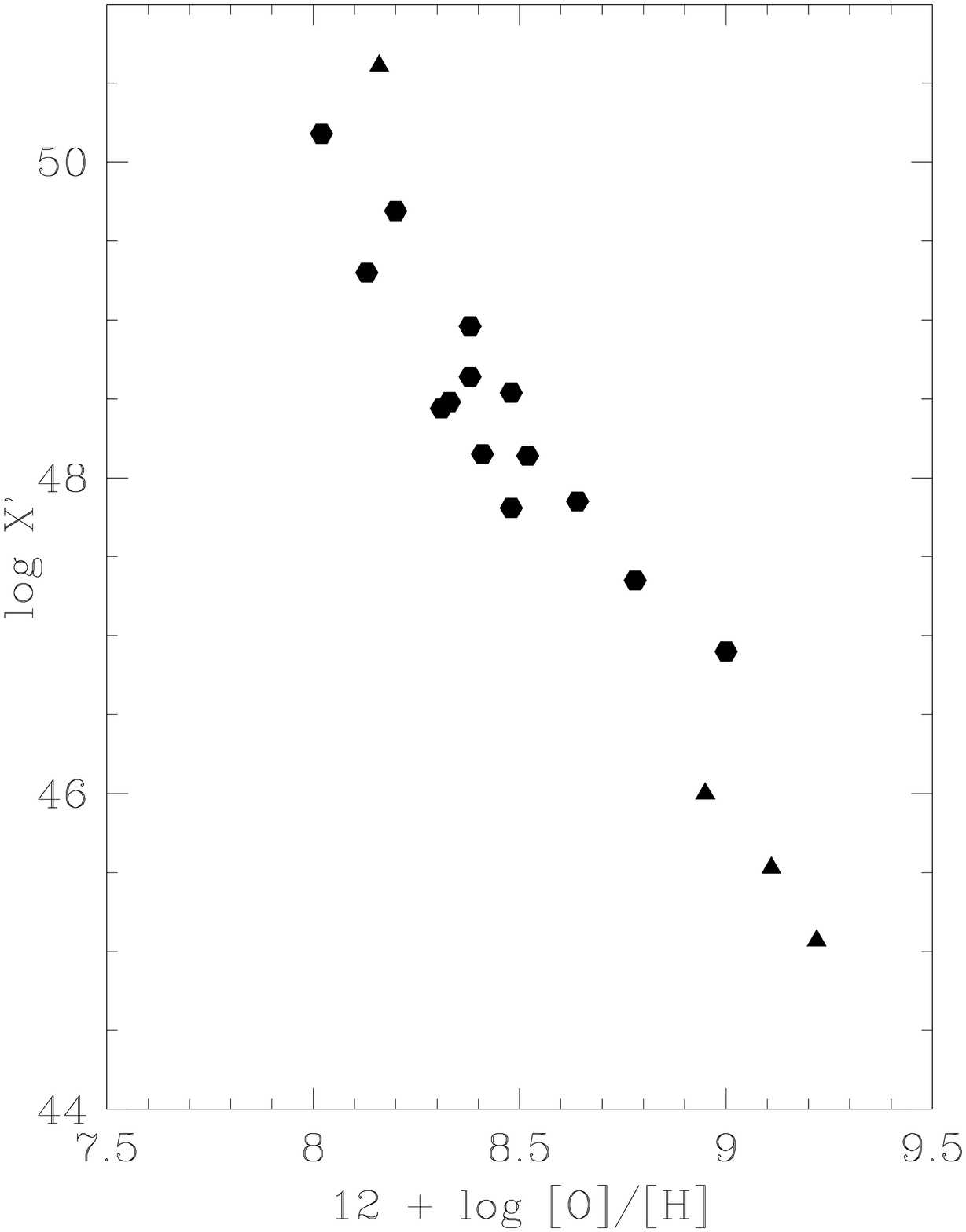,width=6.5cm}
\end{picture}\par
\caption{Dependence of $X'$ on metallicity. For definition and units of $X'$,
see Is97. Filled hexagons: points taken from Is97; filled triangles:
additional points (see text).}
\end{minipage}
\hfill
\begin{minipage}[t]{6.5cm}
\begin{picture}(6.5,9)
\psfig{figure=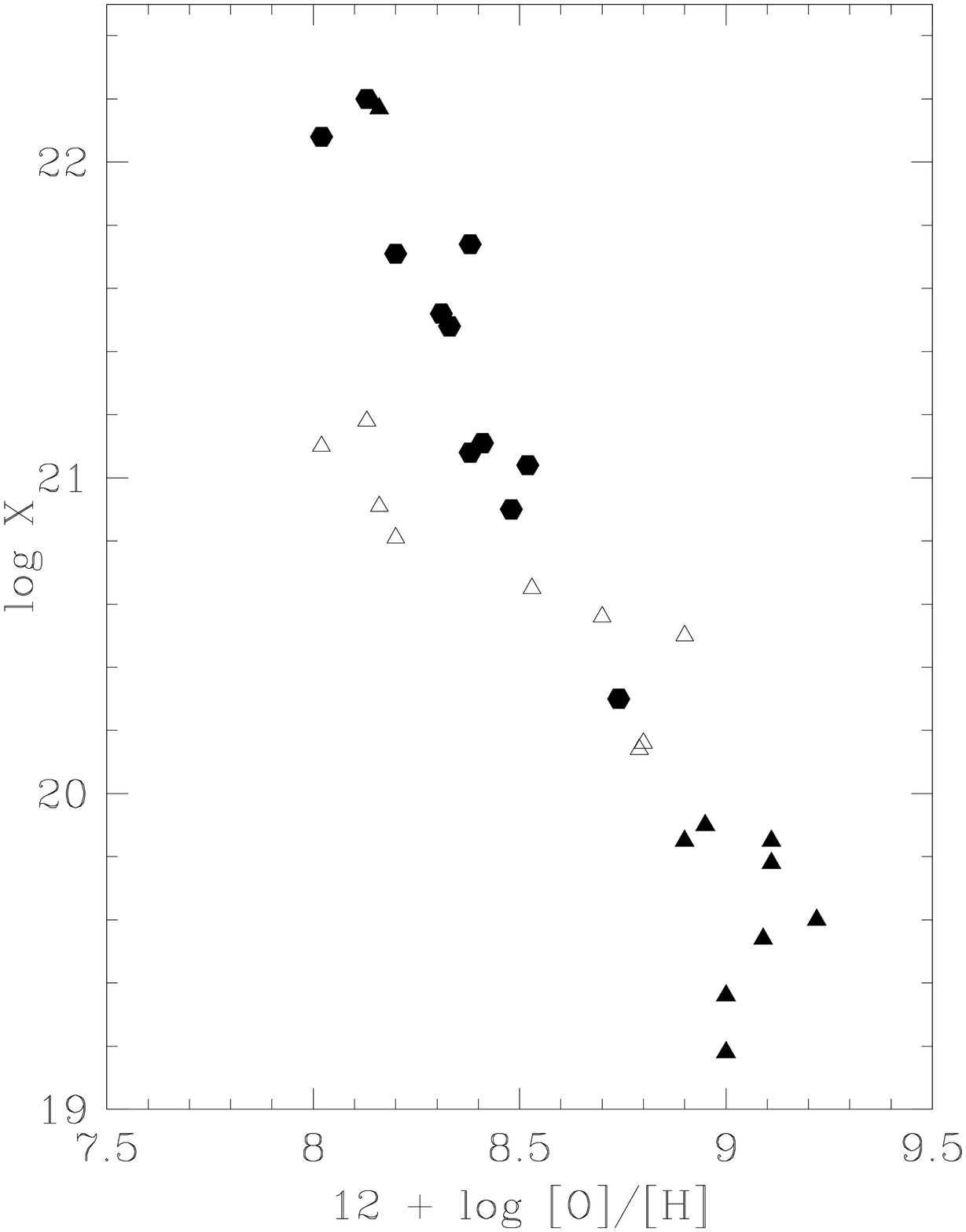,width=6.5cm}
\end{picture}\par
\caption{Dependence of conversion factor $X$ (in mol $\h2$ $\cm2$/$\kkms$)
on metallicity. Open triangles: points from millimeter array observations
(see text); otherwise as Fig.1.}
\end{minipage}
\end{figure}

In Fig. 1, radiation-corrected values $X'$ = $X/\sigma$ are plotted against
metallicity [O]/[H]. In Fig. 2, values of $X$ are plotted in in its usual
form. Figs. 1 and 2 yield the relations:
\begin{equation}
 log\,X'\,=\,log\,X/\sigma\,=\,-\,4\,log\,([O]/[H])\,+\,33.9
\end{equation}

and
\begin{equation}
 log\,X\,=\,-2.5\,log\,([O]/[H])\,+\,12.2
 \label{xplot}
\end{equation}

With a larger sample size, these results differ only slightly from 
those published by Is97. The points representing high-metallicity regions 
in NGC~7331 extend rather well along the relation defined by the 
low-metallicity dwarfs, as do those representing the galaxy centers 
with a larger scatter. Both correlations are highly significant.
Thus, {\it eqn. (3.2) should in general be used to convert CO intensities 
observed in large beams to obtain $\h2$ column densities} within a factor 
of about two. Note that the result may greatly differ from that 
obtained by applying `standard' Milky Way conversion factors (i.e.
lower by a factor of 4--10 for high-metallicity galaxy centers and higher
by a factor of 10--100 for low-metallicity irregular dwarf galaxies).

In Fig. 2, we have also included $X$ values derived by virial theorem 
application to CO clouds mapped with interferometer arrays taken from 
Wilson (1995 -- replacing her M~31 and M~33 metallicities by those 
from Garnett et al. 1999), Taylor $\&$ Wilson (1998) and Taylor et al. 
(1999). These points define a different dependence of 
$X$ on metallicity, with a much shallower slope of only -1.0. Generally,
these $X$ values are much lower than those in Is97.

\section{Discussion}

A {\it steep} dependence of $X$ on metallicity can be understood within 
the context of photon-dominated regions (PDRs). In weak radiation fields 
and at high metallicities, neither $\h2$ or CO suffers much from 
photo-dissociation, and the CO volume will fill practically the whole 
$\h2$ volume. However, when radiation fields become intense, CO 
photo-dissociates more rapidly than $\h2$ because it is less strongly 
selfshielding. Thus, the projected CO emitting projected area will shrink 
and no longer fill that of $\h2$. The observed CO intensity, proportional
to the shrinking emitting area, then requires use of a {\it higher} $X$ 
factor to obtain the correct, essentially unchanged $\h2$ mass. 
We have found that at constant metallicity, $X$ must be increased
linearly with radiation field intensity.

We may somewhat quantitatively estimate the effects of metallicity on 
CO (self)shielding and thereby on $X$. From Garnett et al. (1999) we find 
that over the range covered by Figs. 1 and 2, log [C]/[H] $\propto$ 1.7 
log [O]/[H]. Thus, CO abundances drop significantly more rapidly than 
metallicity [O]/[H], as do dust abundances given by $M_{dust}/M_{gas}$ 
$\propto$ 2 log [O]/[H] (Lisenfeld $\&$ Ferrara 1998). Thus, a ten times 
lower metallicity (cf. Figs 1 and 2) implies a [CO]/[$\h2$] ratio lower 
by a factor of 50, and less CO shielding by a factor 
of 5000! The precise effect on $X$ depends on the nature of the cloud
ensemble, but at lower metallicities PDR effects very quickly increase 
in magnitude. In a standard $\h2$ column, there is less CO to begin with, 
and this smaller amount is even less capable of resisting further erosion 
by photodissociation. With decreasing metallicity, CO is losing both its 
selfshielding and its dustshielding, so that even modestly strong 
radiation fields completely dissociate extended but relatively low-density 
diffuse CO gas, leaving only embedded smaller higher-density CO clumps 
intact. As CO intensities primarily sample emitting surface area, the loss 
of extended diffuse CO strongly reduces them, even when actual CO mass loss
is still modest. Further metallicity decreases cause further erosion 
and molecular clumps of ever higher column density lose their CO gas. 
CO is thus occupying an ever-smaller fraction of the $\h2$ cloud which 
still fills most of the PDR. Its destruction releases a large amount of atomic 
carbon which is ionized and forms a large and bright cloud of CII filling the 
entire PDR. This and the expected anticorrelation between CO and CII 
intensities is indeed observed in the Magellanic Clouds and in IC~10 (Is97; 
Israel et al. 1996; Madden et al. 1997; Bolatto et al. 2000). As the strongly 
selfshielding $\h2$ is still filling most of the PDR (cf. Maloney $\&$ Black 
1988), the appropriate value of $X$ becomes ever higher. In the extreme case 
of total CO dissociation, any amount of $\h2$ left defines an infinitely large 
value of $X$! 

In contrast, use of e.g. interferometer maps to find resolved CO clouds for 
the determination of $X$ introduces a strong bias in low-metallicity
environments. In the PDR, only those subregions are selected which have 
most succesfully resisted CO erosion, with CO still filling a relatively 
large fraction of the local $\h2$ volume. The relatively low $X$ values 
thus derived, although appropriate for the selected PDR subregions, are
not at all valid for the remaining PDR volume where CO has been much 
weakened or has disappeared; the PDR has a much higher overall $X$ value
than the selected subregion. It is because of this bias that the 
array-derived points in Fig. 2 are much lower than the large-beam points
and exhibit a much weaker dependence on metallicity. Incidentally, it 
also explains the suggested dependence of $X$ on observing beamsize 
(Rubio et al. 1993).

\end{document}